\pdfoutput=1
\documentclass[11pt]{article}
\usepackage[margin=1.4in]{geometry}

\usepackage{amsmath,amssymb}
\usepackage{amsthm}
\usepackage{hyperref}
\usepackage{url}
\usepackage{listings}
\usepackage{xcolor}

\newtheorem{theorem}{Theorem}[section]
\newtheorem{conjecture}[theorem]{Conjecture}

\definecolor{dkgreen}{rgb}{0,0.6,0}
\definecolor{ltblue}{rgb}{0,0.4,0.4}
\definecolor{dkblue}{rgb}{0,0,0.8}
\definecolor{dkviolet}{rgb}{0.3,0,0.5}

\def\({\begin{eqnarray*}}
\def\){\end{eqnarray*}}
\newenvironment{eqn}{\begin{eqnarray*}}{\end{eqnarray*}}
\newenvironment{program}{\[\begin{array}l}{\end{array}\]}

\newcommand{\Rule}[4]{
\makebox{  
$\displaystyle
\frac{\begin{array}{l}#2\\\end{array}}
{\begin{array}{l}#3\\\end{array}}$
 #4}}

\def\bnf{\ |\ }

\def\l{\lambda}
\def\rew{\longrightarrow}
\def\Ra{\Rightarrow}

\def\l{\lambda}

\def\rew{\Ra}

\def\l{\lambda}
\def\rew{\longrightarrow}
\def\Ra{\Rightarrow}

\begin{document}


\title{Simple Types for Polymorphic Functions}
\author{
  Barry Jay \\
  \texttt{Barry.Jay8@gmail.com}
  \and
  Johannes Bader (Jane Street) \\
  \texttt{johannes-bader@outlook.com}
}
\date{}

\maketitle

\begin{abstract}
This paper introduces a simple type system for combinatory logic in which combinators have at most one type, whose polymorphism is revealed by application. The combinatory types exactly describe the structure of their values, which may be hidden by abstract types, such as list types and function types. Even without any quantified types, it supports polymorphism beyond that of the Hindley-Milner type system that underpins functional programming, and an effective type inference algorithm. Also, the simplicity of the formalism should make other static program analyses easier. 
\end{abstract}

\section{Introduction}
\label{sec:intro}

Combinatory logic was created by Sch\"onfinkel \cite{Schoenfinkel24} to eliminate the quantifiers from predicate calculus. In modern terms, the representation  $\l x. x$ of a proof that $p$ implies $p$ can be replaced by the identity combinator $I = SKK$. However, the quantifiers reappear when the proposition above is represented by the quantified type $\forall X. X\to X$. This paper introduces {\em combinatory types}, without type variables,  their quantifiers, or even function types, in which $I$ has type $S_2 K_0 K_0$. Every term has at most one type, whose polymorphism is instantiated on application to a term. This improves on System~F, which specialises a term by applying it to a type.  The combinatory type system supports  polymorphism beyond that of the Hindley-Milner type system  (HM) \cite{mil78} while still being able to infer types. The simplicity of the combinatory types should provide a good basis for other static program analyses.

While the core system can type polymorphic functions, polymorphic data requires something more. In HM, constants are introduced in an ad hoc manner. For example, a type declaration can be used to introduce a new type form ${\bf List}~U$ of lists and new term constants {\bf nil} and {\bf cons} for constructing them.
Here, too, type declarations will be used to introduce new type forms, namely the {\em abstract types}. However, their constructors and destructors will be represented by pre-existing combinators built from $S$ and $K$. This is possible because combinatory logic can {\em tag} a function $f$ by a label  $t$. Although tagging does not change the functionality of a term, it allows typing to distinguish between isomorphic types.

Surprisingly, there are also abstract types of functions and recursive functions. By abstracting away internal structure, recursive functions can preserve their type across iterations. 

The simplicity of the {\em abstract combinatory type system} makes it especially convenient for static program analysis. For example, the layers of abstract types could be used to support different layers of abstract interpretation \cite{AH87}. In particular, since the programs are already normal forms, which is possible in combinatory logic \cite{Jay:2018a} but not in $\l$-calculus, there is no need to encode them as syntax trees before analysis. To illustrate the possibilities, we develop an effective type inference algorithm. That is, the algorithm finds the unique type if it has one, fails only if the combinator does not have a type, but may not terminate in pathological cases. This algorithm types all of the usual polymorphic programs that motivated the Damos-Milner algorithm \cite{10.1145/582153.582176}. The system also types structures containing polymorphic data that are beyond the traditional algorithm. Interestingly, it does this by evaluating the {\em type application function}, used to type applications of combinators, which is easier than type unification. Future work is expected to apply these techniques to {\em typed tree calculus} \cite{10.1145/3704253.3706138}, where program analyses, such as self-interpretation, are internalised as first-class programs. Then the types themselves could be internalised.

The paper develops type systems of increasing expressive power by adding abstract types. These developments are summarised in the abstract combinatory type system of Section~\ref{sec:summary}. There are formal proofs, in Rocq  \cite{Jay-combinatory-types-proofs}, that 
\begin{itemize}
\item
reduction preserves typing
\item
every combinator has at most one type
\item
there is an effective type inference algorithm.
\end{itemize}
Also, the type inference algorithm has been implemented in OCaml and  tested on examples from the paper \cite{BaderCombTypes2025}. 

The structure of the paper is as follows. 
Section~\ref{sec:intro} is the introduction. 
Section~\ref{sec:terms} discusses combinators.
Section~\ref{sec:functional_types} adds combinatory types to  the Hindley-Milner type system.
Section~\ref{sec:types} considers the combinatory types in isolation. 
Section~\ref{sec:abstraction} introduces the abstract types.
Section~\ref{sec:funty} introduces function types and recursion types. 
Section~\ref{sec:arithmetic} shows Turing completeness. 
Section~\ref{sec:lists} introduces lists and folding. 
Section~\ref{sec:summary} presents the summary system and two main theorems.
Section~\ref{sec:infer} develops type inference and its implementation in OCaml.
Section~\ref{sec:related} discusses related work. 
Section~\ref{sec:future} considers future work. 
Section~\ref{sec:conclusions} draws conclusions.

\section{Terms}
\label{sec:terms}

\subsection{Combinators}

The {\em combinators} are given by the BNF
\[
M,N := S \bnf K \bnf MN\; .
\]
Application is left-associative. Their reduction rules are 
\begin{eqn}
SM N P &\rew& MP(NP) \\
KMN &\rew& M \; .
\end{eqn}%

Although the reduction rules can be applied in any order within a combinator (it is a rewriting system), fine control over reduction is also supported. For example, the combinatory form 
\[
{\bf wait} \{M,N\} = S(S(KM)(KN) )I
\]
keeps $M$ and $N$ apart until an argument $P$ is given at which time ${\bf wait}\{M,N\} P$ reduces to $MNP$. Thus ${\bf wait}\{M,N\}$ is a normal form if $M$ and $N$ are. 
Similarly,
\[
{\bf wait2}\{M,N,P\} = S(S(S(KM)(KN))(KP))I 
\]
reduces to $MNPQ$ on application to $Q$. 

Combinators will also represent the constructors for abstract types in Section~\ref{sec:abstraction}. Such constructors must carry both the functionality required to eliminate a data structure, given by some combinator $M$  and also some {\em intensionality}, i.e.\ type information. For example, a representation of the truth value ${\bf tt}$ must be able to project from a pair of branches to a conditional, and also determine that its type is {\bf Bool} only. More generally, the combinator $S(S(KK)M)N$ {\em tags} a combinator $M$ by another combinator $N$ while retaining the functionality of $M$ since 
$S(S(KK)M)N P$ reduces to $MP$
by 
\begin{eqn}
S(S(KK)M)N P 
&\rew& S(KK)MP(NP) \\
&\rew& KKP(MP)(NP) \\
&\rew& K(MP)(NP) \\
&\rew& MP \; .
\end{eqn}%
This is good enough for reduction, but there are two challenges for typing. First, the reduction sequence above introduces the combinator $NP$ which must be typed, even though it is to be soon discarded. 
This is easily handled by replacing $N$ with $KN$. The second challenge is to ensure that combinators have unique types by separating these constructors  from the $\l$-abstractions introduced in Section~\ref{sec:lambda}. This is achieved by making the tag incorporate a combinator that will not be a $\l$-abstraction, namely $S(KK)(KK)$.
Define
\begin{eqn}
{\bf tag} &=& S(S(KK)(KK)) \\
{\bf  tagged}\{f,t\}&=& S(S(KK)f)({\bf tag}~(Kt)) \; .
\end{eqn}%
Define the {\em tagged terms} to be those of the form ${\bf tagged}\{f,t\}$. Their functionality is ensured by:

\begin{theorem}[tagged\_red] 
${\bf tagged}\{f,t\}~u \rew f~u$  for all terms $f, t$ and $u$. 
\end{theorem}

\subsection{Lambda Abstractions}
\label{sec:lambda}

In order to define lambda-abstraction, the system needs {\em term variables} $x,y,z,\ldots$. Define the {\em terms} by the BNF
\[
p,q,r,s,t,u,v ::= S \bnf K \bnf t u \bnf x \; .
\]
There are several ways to support lambda-abstraction on terms. The simplest approach is the {\em bracket abstraction} given by 
\begin{program}
\; [x ] t = \\
{\bf match}~t~{\bf with} \\
\bnf y \Ra \mbox{\bf if}~y=x~{\bf then}~I~{\bf else}~Ky \\
\bnf S\Ra KS \\
\bnf K \Ra KK \\
\bnf t_1~t_2 \Ra S([x] t_1)([ x] t_2)\; .
\end{program}%
Bracket abstraction (and star-abstraction below) are presumed to bind as far to the right as possible. 
For example, 
\begin{eqn}
[x] KSS x 
&=& S([x] KSS) [x] x  \\
&=& S(S([x] KS) ([x] S)) [x] x \\
&=& S(S(S ([x] K) ([x]S)) ([x] S)) [x] x \\
&=& S(S(S(KK)(KS))(KS))I \; .
\end{eqn}%

This example illustrates two things about bracket abstractions: they are always normal forms; and  they are quite verbose. Thus, since every computable function can be represented by a bracket abstraction, it follows that every computable function can be represented by a closed normal form \cite{Jay:2018a}.
For this reason, we can define the {\em programs} or {\em values} to be the combinators in normal form. We do not need to consider non-terminating computations as programs. 
This normality makes program analysis much more direct. 

On the other hand, the verbosity of bracket abstraction is a burden.  Although there are many possible ways to optimise bracket abstraction, there is one which is most convenient for our purposes. Define {\em star abstraction} by 
\begin{program}
\l x .t = \\
{\bf match}~t~{\bf with} \\
\bnf y \Ra \mbox{\bf if}~y=x~{\bf then}~I~{\bf else}~Ky \\
\bnf S\Ra KS \\
\bnf K \Ra KK \\
\bnf t_1~t_2 \Ra {\bf if}~x\in {\bf fv}(t_1t_2)~{\bf then}~ S(\l x. t_1)(\l x. t_2)~{\bf else}~K(t_1t_2) \; .
\end{program}%
where ${\bf fv}(t)$ is defined to be the collection of {\em free variables} of $t$ as usual.
Now  
\begin{eqn}
\l x. KSS x
&=& S(\l x. KSS)(\l x. x) \\
&=& S(K(KSS))I 
\end{eqn}%
which is much smaller than the bracket abstraction. It is tempting to go even further, but there are two difficulties. First, optimisations make proofs harder to follow. Second, it is important to maintain the separation of constructors and $\l$-abstractions alluded to in Section~\ref{sec:types}.  In particular, if $\eta$-contraction is used to define $\l x. c~x = c$ whenever $x$ is not free in $c$ then all constructors $c$ would also be abstractions. 

\begin{theorem}[tagged\_not\_star]
$\l x. u \neq {\bf tagged}\{f,t\}$ for all terms $u, f$ and $t$ and variables $x$. 
\end{theorem}


\section{Functional Combinatory Types} 
\label{sec:functional_types}

Our first attempt at typing the combinators is a hybrid type system that adds combinatory types as subtypes of the monotypes of the Hindley-Milner type system.  The terms are those of combinatory logic, with lambda-abstractions represented by bracket abstractions, of the form $[x] t$ and a let-term of the form ${\bf let}~x=u~{\bf in}~t$ being represented by $([x] t)u$. These representations support the expected type derivation rules from HM, so that HM can be viewed as a subsystem. There is also a subsystem which uses only  the combinatory types. 
We show how a type derivation for a program  $p:T$ in the hybrid system   can be transformed  into a type derivation  $p:C$ in the combinatory subsystem followed by a subtyping $C<T$. In this sense, all  typed  programs of HM have principal combinatory types, without using any type schemes. 
This section presumes some familiarity with the mechanics of substitution, etc.\  which won't be required in later sections. 

 The {\em hybrid} type system has  {\em monotypes} $T$ and {\em type schemes} $\tau$ given by  the BNFs
\begin{eqn}
T,U,V &::=& S_0 \bnf S_1 U \bnf S_2 U V \bnf K_0 \bnf K_1 U \bnf X \bnf U\to V \\
\tau &::=& T \bnf \forall X. \tau \; .
\end{eqn}%

The {\em combinatory type forms} are the forms $S_0, S_1 U, S_2 U V, K_0 , K_1 U$. A {\em combinatory type} is one built solely from combinatory forms; they are used to type the combinators in normal form. The {\em type variables} $X$ and {\em function types} $U\to V$ and type schemes are as in HM. 
 
A type scheme $\tau = \forall X_1.\ldots \forall X_n. U$ can be {\em instantiated} to a type $T$ (written $\tau \prec T$) if  $T$ is a substitution $\{U_1/X_1, \ldots U_n/X_n\}U$ of a monotype $U_i$ for $X_i$ in $U$, where substitution is defined in the obvious manner. {\em Type contexts} are sequences of type assignments $x:\tau$ of a type scheme $\tau$ to a term variable $x$. If all of the type schemes are combinatory types then the type context is {\em combinatory}. 
The {\em free type variables} of a type scheme $\tau = \forall X_0.\ldots \forall X_n. T$ are the type variables in $T$ that are not among $X_0,\ldots X_n$. The {\em closure} ${\bf Cl}\{\Gamma,U\}$ of a monotype $U$ in a type context $\Gamma$ quantifies all type variables in $U$ that are not free in $\Gamma$. 
For example, the closure of $X\to Y\to X$ in the context $y:Y$ is $\forall X. X\to Y\to X$. This can be instantiated to $U\to Y\to U$ for any monotype $U$. 

To apply a combinator to an argument requires that it take a function type. This will be managed by a {\em subtyping relation} $U<V$ defined in Figure~\ref{fig:subtyping}. Most of the rules are structural, but those which convert $K_1 U$ and $S_2(U\to V\to W)(U\to V)$ to function types align with the reduction rules of the operators. For example, the subtyping 
\begin{eqn}
S_2 K_0 K_0 &<& S_2 (U\to K_1 U)(U\to K_1U) \\
&<& S_2 (U\to K_1 U \to  U)(U\to K_1 U) \\
&<& U\to U 
\end{eqn}%
yields the expected type of the identity $SKK$. 

\begin{figure}
\[
\begin{array}{rcll}
K_0 &<& K_0 \\
K_0 &<& U\to K_1 U \\
K_1 U_1 &<& K_1 U_2  &\mbox{if $U_1 < U_2$} \\
K_1 U &<& V \to U \\
S_0 &<& S_0 \\
S_0 &<& U\to S_1 U \\
S_1 U_1 &<& S_1 U_2  &\mbox{if $U_1 < U_2$} \\
S_1 U_1 &<& U_2\to S_2 U_1 U_2 \\
S_2 U_1 V_1 &<& S_2 U_2 V_2 &\mbox{if $U_1 < U_2$ and $V_1 < V_2$} \\
S_2(U\to V\to W)(U\to V) &<& U\to W \\
U_1 \to V_1 &<& U_2 \to V_2 &\mbox{if $U_2 < U_1$ and $V_1 < V_2$} \\
U &<& W &\mbox{if $U<V$ and $V<W$}
\end{array}
\]
\caption{Subtyping}
\label{fig:subtyping}
\end{figure}

The type derivation rules are given in Figure~\ref{fig:hybrid}, on the understanding that the usual type derivation rule for $\l$-abstractions becomes a lemma about bracket-abstractions, and ${\bf let}~x=u~{\bf in}~t$ is syntactic sugar for $([x]t)u$. Thus they support the usual rules of HM plus rules for typing the operators $S$ and $K$ and a {\em subsumption rule} for exploiting subtyping. For example, the identity combinator $SKK$ can be typed by 
\[
\Rule{}
{
\Rule{}{
\Rule{}{
\Rule{}{\vdash S : S_0}
{\vdash S : K_0 \to K_0 \to S_2 K_0 K_0 }{}
\quad 
\begin{array}{l}
\ \\
\vdash K : K_0
\end{array}
}
{\vdash SK : K_0 \to S_2 K_0 K_0}{}
\quad 
\begin{array}{l} \\
\vdash K : K_0 
\end{array}
}
{\vdash SKK : S_2 K_0 K_0}{}
}
{\vdash SKK : U\to U}{}
\]
once the side-conditions on subtyping have been checked. 
\begin{figure}
\[
\begin{array}{ccc}
\Rule{}{}{\Gamma\vdash S: S_0}{} 
&&
\Rule{}{\Gamma\vdash M : U\to V \quad \Gamma\vdash N : U}{\Gamma\vdash MN : V}{} \\
\ \\
\Rule{}{}{\Gamma\vdash K : K_0}{}
&\hspace*{1cm}& 
\Rule{}{}{\Gamma\vdash x :T}{$x:\tau\in\Gamma$ and $\tau \prec T$} \\
 \ \\
\Rule{}{\Gamma\vdash t:T}{\Gamma\vdash t:T'}{$T <T'$}
 &&
\Rule{}{\Gamma \vdash u:U \quad x:{\bf Cl}\{\Gamma,U\}, \Gamma \vdash t:T}
{\Gamma\vdash {\bf let}~x=u~{\bf in}~t : T}{} \\ \ \\ 
\end{array}
\]
\caption{Type Derivation for Hybrid Terms}
\label{fig:hybrid}
\end{figure}

The hybrid system supports polymorphism in two different ways, which can be illustrated by the self-application of the polymorphic identity ${\bf let}~f=I~{\bf in}~f~f: U\to U$.  Using the Hindley-Milner approach, the combinator $I$ has type $X\to X$ whose closure is $\forall X. X\to X$ and the body $f~f$ is typed by 
\[
\Rule{}
{
f: \forall X. X \to X \vdash f: (U\to U)\to (U\to U)
\quad 
f: \forall X. X \to X \vdash f: U\to U
}
{f: \forall X. X \to X \vdash f~f : U\to U}{.}
\]

Using combinatory types, we can replace the original let-term by $(SII)I$ which is typed by 
\[
\Rule{}{
\Rule{}{
\Rule{}{
\vdash I : I_0 \to I_0 \to I_0
\quad \vdash I : I_0\to I_0
}
{\vdash SII: I_0\to I_0}{}
\quad 
\Rule{}{}{\vdash I : I_0}{}
}
{\vdash (SII)I: I_0}{}
}
{\vdash (SII)I:  U\to U}{$I_0 < U\to U$}
\]
where $I_0 = S_2 K_0 K_0$.
This eliminates the type schemes, and delayed introduction of function types and type variables until the subtyping $I_0 < U\to U$. We believe this illustrates a general result, expressed as the following conjecture. 

\begin{conjecture}
If there is a type derivation  $\Gamma\vdash t:T$ in the hybrid system where $\Gamma$ is combinatory then there is a combinatory type $C$ and a derivation $\Gamma \vdash  t  :C$ in the combinatory subsystem such that $C < T$. 
\end{conjecture}

The case analyses for the proof have been written out, but we have yet to find the right induction principle to combine them. 

In particular, if $t$ is a typed program of HM then it can be typed here in an empty context, which is automatically combinatory, suggesting that the program has a principal type in the combinatory subsystem. 

This approach eliminates the type variables and their quantification, but still relies on function types to apply combinators. For example, the application $SK$ uses the subtyping $S_0 < K_0 \to S_1 K_0$. However, by replacing subtyping with {\em type applications} such as $S_0(K_0) = S_1 K_0$ it will be enough to support the combinatory types alone.

\section{Combinatory Types}
\label{sec:types}

The simplest type system for the combinators is the {\em combinatory type system} that uses the {\em combinatory types} only. They are defined by the BNF
\[
T,U,V := S_0 \bnf S_1 U \bnf S_2 U V \bnf K_0 \bnf K_1 U 
\]
without type variables or function types. Instead of subtyping rules of the form $T< U\to V$, applications will be typed by {\em type applications} $T(U) = V$. This is a partial function, here defined by 
\begin{eqn}
S_0 (U) &=& S_1 U \\
S_1U(V) &=& S_2 U V \\
S_2T_1T_2(U) &=& T_1 (U)(T_2(U))  \\
K_0 (U) &=& K_1 U \\
K_1U(V) &=& U \; .
\end{eqn}%
The rules for applying $S_2$ and $K_1$ reflect the reduction rules for $S$ and $K$. The other rules are structural. At a glance, type application may appear to be a total function, but the recursive call in $S_2T_1T_2(U)$ will terminate only if the corresponding reduction does. For example, $(SII)(SII)$ reduces to itself, and so $(S_2 I_0 I_0)(S_2 I_0 I_0)$ does not have a value.

\begin{figure}
\[
\Rule{}{}{S: S_0}{} 
\quad 
\Rule{}{}{K : K_0}{}
\quad 
\Rule{}{M : T \quad N : U}{MN : V}{$T (U) = V$.}
\]
\caption{Type Derivation for Combinators}
\label{fig:combinatory}
\end{figure}

The {\em type derivation rules} are given in Figure~\ref{fig:combinatory}. Since type application is a functional relation, each combinator has at most one type, which is, perforce, its principal type, whose specialisation is triggered by its application. For example, if $SKK$ of type $I_0 = S_2 K_0 K_0$ is applied to some combinator $M:T$  then the type application is $(S_2K_0K_0)(T) = K_0(T)(K_0(T)) = T$.
For now, each type has exactly one value, which is not very interesting, but this will change once we add abstract types in Section~\ref{sec:abstraction}.  In anticipation of this, let us establish some notation. 

In this setting, each program $p$ has its corresponding type $|p|$ defined in the obvious manner, by 
\begin{eqn}
|S| &=& S_0 \\
|Sp| &=& S_1 |p| \\
|Spq| &=& S_2~|p|~|q| \\
|K| &=& K_0 \\
|Kp| &=& K_1 |p| \; .
\end{eqn}%

Each tagged term ${\bf tagged}\{f,t\}$ where $f:F$ and $t:T$ has a corresponding {\em tagged type}. It  is defined by 
\[
{\bf tagged\_ty}\{F,T\} = S_2(S_2(K_1 K_0) F)(S_2 (S_2(K_1 K_0)(K_1 K_0)) (K_1 T)) 
\]
which can be viewed as the result of a type application ${\bf tagged\_tyl}\{F\}({\bf tag\_ty}\{T\})$ where 
\begin{eqn}
{\bf tagged\_tyl}\{F\} &=& S_1(S_2(K_1 K_0) F) \\
{\bf tag\_ty}\{T\} &=& S_2 (S_2(K_1 K_0)(K_1 K_0)) (K_1 T) \; .
\end{eqn}%
Later, this application will be reserved for producing abstract types.

\begin{figure}
\[
\begin{array}{ccc}
\Rule{}{}{\Gamma\vdash S: S_0}{} 
&&
\Rule{}{\Gamma\vdash M : T \quad \Gamma\vdash N : U}{\Gamma\vdash MN : V}{$T (U) = V.$}
\ \ \\
\Rule{}{}{\Gamma\vdash K : K_0}{}
&\hspace*{1cm}& 
\Rule{}{}{\Gamma\vdash x :T}{$x:T\in\Gamma$}
\end{array}
\]
\caption{Type Derivation for Terms}
\label{fig:terms}
\end{figure}

When term variables are admitted, then type derivation requires a context. Define a {\em term context} $\Gamma$ to be a list of type assignments $x:T$ of types $T$ to variables $x$ in the usual way. We may write $x:T\in\Gamma$ if $x:T$ is in $\Gamma$. The corresponding type derivation rules are given in Figure~\ref{fig:terms}.

\section{Abstract Types}
\label{sec:abstraction}

To be useful, the Hindley-Milner type system requires some data types as well as function types, as their representations in System~F are not type schemes in HM. Commonly, these data types are algebraic. For example, list types have been traditionally introduced by a declaration that is something like
\[
{\bf List}~U = {\bf nil} \bnf {\bf cons}~{\bf of}~U \ast  {\bf List}~U\; .
\]
[This particular declaration will be modified slightly in Section~\ref{sec:lists}, in ways that do not impact on this discussion.] The declaration introduces polymorphic constructors ${\bf nil}: {\bf List}~U$ and ${\bf cons}: U\ast {\bf List}~U \to {\bf List}~U$ which can be eliminated by pattern-matching. 

 Here, these declarations will be interpreted a little differently. First, the declared types and untyped terms will already be supported. For example, \\${\bf List}~U =  {\bf Abs}_1 \{S_2 (S_1 S_0)S_0\} U$ where the type $S_2 (S_1 S_0)S_0$ acts as a label for lists. The particular choice of label is immaterial as long  as the various abstract types are kept apart. Also, the constructors  and destructors will be represented by $SK$-combinators. The constructors are given by tagging their traditional representation as functions with a label that determines their abstract type. The destructors unpack the usual pattern-matching. The actual impact of the declaration is to introduce new type application rules for introducing and eliminating the abstract types, which allow the constructors and destructors to be appropriately typed.  For example, eliminating lists requires a pair of functions, to handle nil and cons, so we add
 \[
( {\bf List}~U)(V\ast T) = V \quad \mbox{ if $T(U\ast {\bf List}~U) = V$} \; 
\]

Although this machinery has not yet been automated, the development will rely on type declarations as much as possible, leaving the details to the summary type system of Section~\ref{sec:summary} which is formalised in the Rocq presentation \cite{Jay-combinatory-types-proofs}.

In general, abstract types may take any number of type arguments but here, for convenience, there will be at most two. So the {\em abstract combinatory types} are given by 
\begin{eqn}
T,U,V &::=&S_0 \bnf S_1 U \bnf S_2 U V \bnf K_0 \bnf K_1 U \bnf \\
 && {\bf Abs}_0\{ T\} \bnf    {\bf Abs}_1 \{T\} ~U \bnf    {\bf Abs}_2\{ T\}~ U~ V 
\end{eqn}%
where the type $T$ in $\{T\}$ serves to label the type. Note that function types are not primitive here, but may be introduced as abstract types in Section~\ref{sec:funty}. When abstract types are supported then the result is an {\em abstract combinatory type system}. Such systems may also be called {\em combinatory type systems} if no confusion results. 

The values of abstract types will be tagged terms, as introduced in Section~\ref{sec:types}.  Since type application is to be functional, it follows that tagged terms of abstract type cannot also be of combinatory type. The simplest resolution is to restrict the existing rule for applying types of the form $S_1 U$ to become
\[
S_1 U(V) = S_2 U V \quad \mbox{if $S_2 U V$ is not a tagged type.}
\]
It follows that tagged terms do not have a type unless a specific type application rule is introduced for them. 
In this way, new abstract types can be added without further modification to the rules for combinatory types. 

\subsection{Product Types} 

Since product types will be used to create tuples of constructor arguments, they must be introduced directly. 
Define $U\ast V = {\bf Abs}_2 \{S_1S_0\} ~U~V$ with pairing and projections given by 
\begin{eqn}
{\bf pair} &=& \l x. \l y. {\bf tagged}\{\l f. f x y, {\bf product\_tag} \}\\
{\bf fst}&=& \l p. p (\l x. \l y. x) \\
{\bf snd}&=& \l p. p (\l x. \l y. y) \; .
\end{eqn}%
where {\bf product\_tag} is some program that has been chosen to indicate a product type, namely $SS$. 
In turn, these definitions guide the choice of new type application rules
\begin{eqn}
 {\bf tagged\_tyl}\{S_2 (S_2 I_0 (K_1 U))(K_1 V)\}~{\bf tag\_ty}\{ | {\bf product\_tag}| \}&=& U\ast V \\
 (U\ast V)(T) &=& T(U) (V) \; .
\end{eqn}%

Thus, if $u:U$ and $v:V$ then 
\begin{eqn}
{\bf pair}~u~v &:& U\ast V \\
{\bf fst}~({\bf pair}~u~v) &\rew& {\bf pair}~u~v~K \rew K~u~v \rew u \\
{\bf snd}~({\bf pair}~u~v) &\rew& {\bf pair}~u~v~(KI) \rew KI~u~v \rew v 
\end{eqn}%
as expected.  Note how the reductions of ${\bf pair}~u~v~K$ and ${\bf pair}~u~v~(KI)$ are mirrored in the type application rule $(U\ast V)T = T(U)(V)$.

 \subsection{Booleans}
 
 Booleans are declared by 
 \[
 {\bf Bool} = {\bf tt} \bnf {\bf ff} \; .
 \]
 Detailed interpretations of this type and terms are in Section~\ref{sec:summary}. 
  It follows that 
  \begin{eqn}
  {\bf cond}~{\bf tt}~u~v &\rew& {\bf tt}~({\bf pair}~u~v) \rew {\bf fst}~({\bf pair}~u~v) \rew u \\
  {\bf cond}~{\bf ff}~u~v &\rew& {\bf ff}~({\bf pair}~u~v) \rew {\bf snd}~({\bf pair}~u~v) \rew v 
    \end{eqn}%
as expected. The actual type of {\bf cond} is quite complicated. To see this, unpack the definition of {\bf cond} to get 
 \begin{eqn}
 {\bf cond} &=& \l b. \l x.\l y. b ({\bf pair}~x~y) \\
 &=& \l b. \l x. (S(Kb)({\bf pair}~x)) \\
 &=& \l b. S(K(S(Kb)))~(S (K~{\bf pair})I) \\
 &=& S(S(KS) (S(KK)(S(KS)I))) (K(S(K~{\bf pair})I) \; . 
 \end{eqn}%
 The full expansion must then unpack {\bf pair} and $I$ but the point is made. The type of {\bf cond} is thus 
 \[
 S_2 (S_2(K_1 S_0)(S_2(K_1 K_0) (S_2(K_1 S_0)I_0))) (K_1 (S_2 (K_1 |{\bf pair}|)I_0))\; .
 \]
 This combinatory type exactly describes the structure of {\bf cond} but obscures its functionality, which might be annoying. However, for any terms $b,x$ and $y$ where $b$ is of type {\bf Bool} and $x$ and $y$ are of type $U$ then ${\bf cond}~b~x~y$ has type $U$. That is, we can deduce the following theorem about a triple type application 
 
 \begin{theorem}[app\_ty\_cond]
$| {\bf cond} |({\bf Bool})(U)(U) = U$ for all types $U$. 
\end{theorem}

From this follows

\begin{theorem}[derive\_cond] There is an implied type derivation rule 
 \[
 \Rule{}
 {\Gamma\vdash b:{\bf Bool} \quad \Gamma\vdash  u: U \quad \Gamma\vdash v : U}
 {\Gamma\vdash {\bf cond}~b~u~v : U}{.}
 \]
 \end{theorem}
 
 When an abstract type of functions is introduced in Section~\ref{sec:funty}, it will be possible to tag {\bf cond} to get a combinator of type ${\bf Bool} \to U\to U \to U$.

\subsection{Comparison to Traditional Approaches} 

In traditional approaches, the introduction of types forces changes to the term language. Let us consider a type {\bf Bool} of booleans in $\l$-calculus, with constructors {\bf tt} and {\bf ff} of type {\bf Bool} and a conditional {\bf cond} of type 
\[
{\bf cond} : {\bf Bool} \to U \to U\to U
\]
for every type $U$. In pure $\l$-calculus we can represent {\bf tt} by $\l x.\l y. x$ and {\bf ff} by $\l x. \l y .y$ and {\bf cond} by $\l b.b$. However, the presence of types introduces complications. 

In simply-typed $\lambda$-calculus, all term variables are annotated with types, which changes the term language. Also, every term has a unique type, which forces us to introduce many types of booleans. For example, given a type $U$, we could define  ${\bf tt} = \l x^U \l y^U. x^U$ etc.\ but then each type $U$ must support its own type ${\bf Bool}^U$ of booleans, which is too much. So it is usual to introduce {\bf tt} and {\bf ff} and {\bf cond} as constants.

System~HM uses even more machinery, including let-terms to handle polymorphism, and type unification to support type inference. 

In System~F, the need for constants (and let-terms) is avoided, but term variables are still annotated with types. In addition, terms may be applied to types, or may be abstracted with respect to type variables. Now we have a unique type for {\bf cond} namely
\[
{\bf cond} : \forall X. {\bf Bool} \to X \to X \to X
\]
which can be instantiated to ${\bf cond}~U : {\bf Bool} \to U\to U\to U$ 
but type inference is not supported. Further, System~F types are not intensional, in that all two-values types are identified. 

Happily, combinatory logic can overcome all of these difficulties. The polymorphism of {\bf fst} and {\bf snd} extends to {\bf tt} and {\bf ff} and {\bf cond} while tagging captures the intensionality.

\subsection{Sum Types}

To represent sum types combinatorially presents  a new challenge. The usual type declaration of ${\bf sum}~U~V$ uses constructors ${\bf inl} : U\to {\bf sum}~U~V$ and ${\bf inr}: V \to {\bf sum}~U~V$ but then ${\bf inl}~u$ would be of type ${\bf sum}~U~V$ for any type $V$ instead of having a unique type. Recall that System~F  eliminates the ambiguity by applying {\bf inl} to type arguments $U$ and $V$. 
Since we are not allowing any changes to the combinatory language, we must convey the type information through dummy terms  $d_U:U$ and $d_V:V$ and use a boolean to decide which projection of the pair is important.  The only disadvantage is that $U$ and $V$ must be inhabited, i.e.\ must be the types of some combinators. Although not a practical difficulty, this requirement complicates the theory. Now {\bf inl} is to act on pairs of the form ${\bf pair}~u~d$ where $u:U$ and $d:V$ is a dummy value. Conversely, {\bf inr} is to act on pairs of the form ${\bf pair}~d~v$ where $v:V$ and $d:U$ is a dummy value. So declare
\[
{\bf sum}~U~V = {\bf mk\_sum}~\mbox{of}~{\bf Bool}\ast U\ast V 
\]
and let $U+V = {\bf sum}~U~V$ and define
\begin{eqn}
{\bf inl} &=& \l p. {\bf mk\_sum}( {\bf pair}~{\bf tt}~p) \\
{\bf inr} &=& \l p. {\bf mk\_sum}( {\bf pair}~{\bf ff}~p) \\
{\bf case} &=& \l q. \l c. {\bf fst}~c~({\bf pair}~({\bf fst}~q ~({\bf fst}~({\bf snd}~c)))
~({\bf snd}~q ~({\bf snd}~({\bf snd}~c))))\; . 
\end{eqn}%
This becomes the usual interface since  if $p: U\ast V$ and $u:U$  and $v : V$ then 
\begin{eqn}
{\bf inl}~p &:& U+V \\
{\bf inr}~p &:& U+V \\
{\bf case}~({\bf pair}~f~g)~({\bf inl}~({\bf pair}~u~d)) &\rew& f~u \\
{\bf case}~({\bf pair}~f~g)~({\bf inr}~({\bf pair}~d~v) &\rew& g~v 
\end{eqn}%
as expected. 
Sum types illustrate the need for dummy values that arises in the definitions of function types and list types.

\section{Recursive Function Types}
\label{sec:funty}

\subsection{Function Types}

Traditionally, function types play a central role in type theory, especially in typed $\l$-calculus. Here, they are another form of abstract type, since their purpose is to hide implementation details, just as the type {\bf Bool} hides the structure of {\bf tt} and {\bf ff}.  Declare function types by 
\[
{\bf Funty}~U~V = {\bf mk\_fun}~\mbox{of}~T\ast U~\mbox{where}~T(U) = V \; .
\]
The where-clause is new syntax that constrains the typing as follows.  The construction ${\bf mk\_fun}~({\bf pair}~t~d)$ is of type ${\bf Funty}~U~V$ only if the dummy value $d$ is of type $U$ and the function $t$ has type $T$ where $T(U) = V$. 
Let $U\to V$ be new syntax for ${\bf Funty}~U~V$. That is, the introduction rule for function types is
conditional on $T(U) = V$ while the elimination rule is 
\[
(U\to V) (U) = V \; .
\]
More naturally, we can combine star-abstraction and {\bf mk\_fun} to get 
\[
{\bf lam}~x~t~d = {\bf mk\_fun}({\bf pair}~(\l x. t)~d)\; .
\]
For example, given $d:U$ we can define 
\begin{eqn}
{\bf cond\_mono}\{d\} &=& {\bf lam}~b~({\bf lam}~x~({\bf lam}~y~(b~({\bf pair}~x~y))~d)~d)~{\bf tt} \\
&:& {\bf Bool} \to U\to U\to U
\end{eqn}%
which makes the conditional be of function type, but no longer polymorphic.

\begin{theorem}[derive\_lam] 
 There is an implied type derivation rule 
  \[
  \Rule{}{x:U,\Gamma\vdash t : V \quad \Gamma\vdash d : U}
  {\Gamma\vdash {\bf lam}~x~t~d : U\to V}{.}
  \]
  \end{theorem}
  
Although function types hide many details they also risk eliminating polymorphism, as when the polymorphic {\bf cond} is replaced by the monomorphic {\bf cond\_mono}. Similar trade-offs can be seen with function composition. Using combinators, the composition of a function $f$ followed by a function $g$ is $S(Kg)f: S_2(K_1 G)F$ if $f:F$ and $g:G$. This has the potential to act on many types. Tagging it by some $K_1 d$ where $d:U$ produces a term of type $U\to V$ provided that $G(F(U)) = V$ but is now monomorphic. 
  
  \subsection{Recursive Functions} 
  
Since the $Y$ combinator defined by $Y f \rew f (Y f)$ will never have a normal form, we consider fixpoint {\em functions} specified by 
\[
Z\{f\} x \rew f (Z\{f\})x
\]
where $Z\{f\}$ waits for an argument, and so will be normal if $f$ is. The functionality of recursion is given by defining {\bf Z\_tag} to be $K$ and defining 
\begin{eqn}
\omega &=&
  \l w. 
    \l f. 
       \l x. f ~{\bf tagged}\{{\bf tagged}\{{\bf wait2}\{w,w,f\}, {\bf Z\_tag}\},Kx\}~x \\
 {\bf Z}\{ f\} &=& {\bf tagged}\{{\bf wait2}\{\omega, \omega,f\}, {\bf Z\_tag}\} \; . 
\end{eqn}%
The simplest way to understand this is perhaps to evaluate it, by 
\begin{eqn}
Z\{f\}u 
&=&  {\bf tagged}\{{\bf wait2}\{\omega, \omega,f\}, {\bf Z\_tag}\} ~u \\
&\rew& {\bf wait2}\{\omega, \omega,f\} ~u \\
&\rew& \omega~\omega~f ~u \\
&\rew&  f ~{\bf tagged}\{{\bf tagged}\{{\bf wait2}\{\omega,\omega,f\}, {\bf Z\_tag}\},Ku\}~u \\
&=& f~{\bf tagged}\{Z\{f\},Ku\}~u \; .
\end{eqn}%
Reduction first eliminates the tag, and then the waiting. The next step simulates $\beta$-reduction of $\omega$, which reconstructs $Z\{f\}$. This reduction sequence proves the following theorem:

\begin{theorem}[Z\_red] 
$Z\{f\}~u \rew f~{\bf tagged}\{Z\{f\}, Ku\}~u$ for all $f$ and $u$. 
\end{theorem}

Although this is all legal, it may not yet be clear why there is so much tagging.  So let us introduce the type applications for recursion types. First define 
\begin{eqn}
{\bf wait2\_ty} \{U, V, W\} &=&  S_2 (S_2 (S_2 (K_1 U) (K_1 V))(K_1W)) I_0 \\
{\bf Zty}\{F\} &=& {\bf wait2\_ty}\{| \omega|, | \omega|, F\}\; .
\end{eqn}%
Then the introduction rule is 
\[
{\bf tagged\_tyl}\{{\bf Zty}\{F\}\}~({\bf tag\_ty}\{|{\bf Z\_tag}|\}) = {\bf Rec}\{F\} 
\]
and the elimination rule is 
\[
\Rule{}{F(V\ast U\to V)(V\ast U) = V}{ {\bf Rec}\{F\}~(V\ast U) \Ra V}{.}
\]
This exposes another curiosity. Why does the elimination rule use $V\ast U$ instead of $U$? As before, this is done to ensure that type application is functional, and may require a dummy argument to force the typing. 
Coming back to the terms, the tagging in $Z\{f\}$ is used to produce something of recursion type. Since $\omega$ must recreate $Z\{f\}$ it requires the inner tagging. Its outer tagging is required because $f~Z\{f\}$ is ill-typed. $Z\{f\}: {\bf Rec}\{F\}$ must be tagged to produce something of type $V\ast U\to V$. 

Note that no type declaration has been provided for ${\bf Rec}\{F\}$. There are two reasons for this. First, it is not clear how to represent a conditional elimination rule. Second, it is not clear how such a declaration might be understood without the explanations and calculations above. 

These fixpoint functions can be used to define the $\mu$-recursive arithmetic functions and show Turing-completeness.

\section{Arithmetic}
\label{sec:arithmetic}

The natural number type is declared by 
\[
{\bf Nat} = {\bf zero} \bnf {\bf successor}~\mbox{of}~{\bf Nat}\; .
\]
Natural numbers can be applied to a pair of a $U$ (for the zero-case) and a function ${\bf Nat}$ to $U$ for the successor case. This is enough to type {\bf isZero} and {\bf predecessor} given by 
\begin{eqn}
{\bf isZero} &=& \l n. n ({\bf pair}~{\bf tt}~ (K~{\bf ff})) \\ 
{\bf predecessor} &=& \l n. n ({\bf pair}~{\bf zero}~I)\; .
\end{eqn}%
Now recursive arithmetic functions can be defined using $Z$.  Define
\begin{eqn}
{\bf primrec0}\{g,h\} &=& Z \{\l z.\l p. {\bf snd}~p~({\bf pair}~g~(\l n. (h~n~(z~({\bf pair}~({\bf fst}~p)~n))))) \} \\
{\bf primrec}\{g,h,u\} &=& {\bf primrec0}\{g~u, h~u\}
\end{eqn}%
on the understanding that neither $z$ nor $p$ is free in either $g$ or $h$. Then we have

\begin{theorem}[primrec\_red\_zero]
For all $g,h,u,v$ and $n$ we have\
\begin{eqn}
{\bf primrec}\{g, h,u\} \;({\bf pair}~v~{\bf zero}) &\rew& gu \\
{\bf primrec}\{g, h,u\} \;({\bf pair}~v~({\bf successor}~n))  &\rew& hun\,{\bf primrec}\{g,h,u\}\,({\bf pair}\,v\,n).
\end{eqn}%
\end{theorem}

\begin{theorem}[derive\_primrec\_app]
  There is an implied type derivation rule 
  \[
  \Rule{}{
  \begin{array}{c}
  \Gamma\vdash g:G \quad \Gamma \vdash h:H \quad \Gamma\vdash u:U \quad \Gamma\vdash p: V \ast {\bf Nat} \\ G(U) = T \quad H(U)({\bf Nat})(T) = T
  \end{array}}
  {\Gamma\vdash {\bf primrec}\{g,h,u\}~p : T}{}
  \]
\end{theorem}

For minimal solutions we have 
\begin{eqn}
{\bf minrec}\{f,u\} = Z\{\l z. \l p. {\bf cond} &&(f~u~({\bf snd}~p))~  ({\bf snd}~p) \\
&&(z~({\bf pair}~({\bf fst}~p)~({\bf successor}~({\bf snd}~p)))) \}
\end{eqn}%
on the understanding that neither $z$ nor $p$ occurs in $f$. It follows that 

\begin{theorem}[minrec\_red] For all terms $f, u$ and $n$ , we have 
\[
\begin{array}{rcll}
{\bf minrec}\{f,u\}~n &\rew& n & \mbox{if $f~u~n \rew {\bf tt}$} \\
{\bf minrec}\{f,u\}~n &\rew& {\bf minrec}\{f,u\}~({\bf successor}~n)\quad & \mbox{if $f~u~n \rew {\bf ff}$} \; .
\end{array}
\]
\end{theorem}

\begin{theorem}[derive\_minrec\_app]
   There is an implied type derivation rule 
  \[
  \Rule{}{ \Gamma\vdash f : F \quad \Gamma\vdash u : U \quad \Gamma\vdash n : {\bf Nat} \quad F(U)({\bf Nat}) = {\bf Bool}}
{    \Gamma\vdash {\bf minrec}\{f, u\}~n : {\bf Nat}}{.}
\]
 \end{theorem}
 
Thus, there is an abstract combinatory type system which is Turing-complete.

\section{Lists}
\label{sec:lists}

Like the inclusions of sum types, the nil list does not automatically determine the type of list being considered, and so must take a dummy parameter. 
So declare list types by 
\[
{\bf List}~U = {\bf nil}~\mbox{of}~U \bnf {\bf cons}~\mbox{of}~U\ast{\bf List}~U\; .
\]
Note that the elimination rule for lists types is conditional, given by 
\[
\Rule{}{T(U\ast {\bf List}~U) =V}{({\bf List}~U)(V\ast T) = V}{.}
\]

Then we can define ${\bf fold\_left}\{f\}$ by 
\[
Z\{\l z. \l p. {\bf snd}~p~({\bf pair}~({\bf fst}~p)~(\l q. z~({\bf pair}~(f~({\bf fst}~p)~({\bf fst}~q))~({\bf snd}~q))))\} 
\]
on the understanding that $z, p$ and $q$ do not occur in $f$. It follows that 

\begin{theorem}[fold\_left\_red] For all terms $f,u,d$ we have
\begin{eqn}
{\bf fold\_left}\{f\}~({\bf pair}~u~({\bf nil}~d)) &\rew& u \\
{\bf fold\_left}\{f\}~({\bf pair}~u~({\bf cons}~({\bf pair}~h~t))) &\rew& {\bf fold\_left}\{f\}~({\bf pair}~(f~u~h)~t)\; .
\end{eqn}
\end{theorem}

\begin{theorem}[derive\_fold\_left] 
   There is an implied type derivation rule 
\[
\Rule{}{\Gamma\vdash f:F \quad \Gamma\vdash p : U\ast {\bf List}~V \quad F(U)(V) = U }
{\Gamma\vdash {\bf fold\_left}\{f\}~p : U}{.}
\]
\end{theorem}

\section{The Summary System}
\label{sec:summary}

For the sake of completeness, all of the abstract types of the paper are brought together in the system of  Figure~\ref{fig:summary}, that uses the following type definitions:
\begin{eqn}
U\ast V &=& {\bf Abs}_2 \{S_1S_0\} U V\\
{\bf Bool} &=& {\bf Abs}_0 \{S_1 K_0\} \\
{\bf Nat} &=&  {\bf Abs}_0 \{S_2K_0K_0\} \\
U+V &=& {\bf Abs}_2 \{S_2 K_0 S_0\}U V\\
U\to V &=&  {\bf Abs}_2 \{K_1 K_0\} U V \\
{\bf Rec }\{F\} &=& {\bf Abs}_1 \{K_0\} F \\
{\bf List}\{U\} &=&  {\bf Abs}_1 \{S_2 (S_1 S_0)S_0\} U\; .
\end{eqn}%

\begin{figure}
\small
{\bf Terms}
$
t,u ::= S \bnf K \bnf t u \bnf x 
$
\[
\omega =
  \l w.
    \l f.
       \l x.
          f~
             {\bf tagged}\{
                {\bf tagged}\{
                   {\bf wait}_2\{w,w,f\}, {\bf Z\_tag}\}, Kx\}~x 
                   \]
                   
{\bf Types}
$T,U,V ::=S_0 \bnf \! S_1 U \bnf S_2 U V \bnf K_0 \bnf K_1 U \bnf {\bf Abs}_0\{ T\} \bnf    {\bf Abs}_1 \{T\} ~U \bnf    {\bf Abs}_2\{ T\}~ U~ V $
\begin{eqn}
{\bf tagged\_tyl}\{F\} &=& S_1 (S_2 (K_1K_0) F) \\
{\bf tag\_ty}\{T\} &=& S_2(S_2(K_1K_0)(K_1K_0)) (K_1 T) \\
{\bf tagged\_ty}\{F,T\} &=& S_2 (S_2 (K_1K_0) F)~({\bf tag\_ty}~T ) \\
{\bf wait2\_ty} \{U, V, W\} &=&  S_2 (S_2 (S_2 (K_1 U) (K_1 V))(K_1W)) I_0 \\
{\bf Zty}\{F\} &=& {\bf wait2\_ty}\{|\omega|, |\omega|, F\}
\end{eqn}%

{\bf Type Application}
\begin{program}
T(V) = \\
\mbox{match}~(T,V)~\mbox{with} \\
\bnf (K_0,V) \Ra K_1V \\
\bnf (K_1 U,V) \Ra U \\
\bnf (S_0,V) \Ra S_1 V \\
\bnf (S_2 U W, V) \Ra U(V)(W(V)) \\
\bnf (S_1 U,V) \Ra S_2 U V \quad\mbox{if $S_2 U V$ is not some ${\bf tagged\_ty}\{F,T\}$ } \\
\bnf (U_1 \ast U_2, V) \Ra V(U_1)(U_2) \\
\bnf ({\bf tagged\_tyl}\{S_2(S_2 I_0 (K_1 U_1)) (K_1 U_2)\},{\bf tag\_ty}\{| {\bf product\_tag}|\}) \Ra U_1\ast U_2 \\
\bnf ({\bf Bool}, U\ast U) \Ra U \\
\bnf ({\bf tagged\_tyl}\{| {\bf fst}|\},{\bf tag\_ty}\{ | {\bf bool\_tag}|\}) \Ra {\bf Bool} \\
\bnf ({\bf tagged\_tyl}\{| {\bf snd}|\}, {\bf tag\_ty}\{ |{\bf bool\_tag}|\}) \Ra {\bf Bool} \\
\bnf ({\bf Nat}, V_1\ast V_2) \Ra V_1 \quad \mbox{if $V_2({\bf Nat}) = V_1$} \\
\bnf ({\bf tagged\_tyl}\{| {\bf fst}|\},{\bf tag\_ty}\{ |{\bf nat\_tag}|\}) \Ra {\bf Nat} \\
\bnf ({\bf tagged\_tyl}\{S_2~| {\bf snd}| ~(K_1{\bf Nat})\},{\bf tag\_ty}\{ |{\bf nat\_tag}|\}) \Ra {\bf Nat} \\
\bnf ({\bf U_1 + U_2}, V) \Ra ({\bf Bool} \ast (U_1 \ast U_2))(V)  \\
\bnf ({\bf tagged\_tyl}\{ {\bf Bool}\ast (U_1 \ast U_2)\},{\bf tag\_ty}\{ |{\bf sum\_tag}|\}) \Ra U_1 + U_2 \\
\bnf ({\bf tagged\_tyl}\{| {\bf snd}|\},{\bf tag\_ty}\{ |{\bf bool\_tag}|\}) \Ra {\bf Bool} \\
\bnf (V\to U,V) \Ra U \\
\bnf ({\bf tagged\_tyl}\{U\},  K_1V_1) \Ra V_1 \to W \quad\mbox{if $U(V_1) = W$} \\ 
\bnf ({\bf Rec}\{F\}, V_1\ast V_2) \Ra V_1 \quad \mbox{if $F(V_1\ast V_2\to V_1)(V_1\ast V_2) = V_1$} \\
\bnf ({\bf tagged\_tyl}\{{\bf Zty}\{F\}\},{\bf tag\_ty}\{ | {\bf Z\_tag}|\}) \Ra {\bf Rec}\{F\} \\
\bnf ({\bf List}\{U\}, V_1\ast V_2) \Ra V_1 \quad \mbox{if $V_2(U\ast {\bf List}\{U\}) = V_1$} \\ 
\bnf ({\bf tagged\_tyl}\{ | {\bf fst}|\}, S_2(S_1 S_0)U) \Ra {\bf List}\{U\} \\
\bnf ({\bf tagged\_tyl}\{ S_2 (S_2 (K_1 | {\bf snd}|)I_0)(K_1(U\ast {\bf List}\{U\})), S_2(S_1 S_0)U)  \Ra {\bf List}\{U\} 
\end{program}

{\bf Type Derivation}
\[
\begin{array}{ccc}
\Rule{}{}{\Gamma\vdash S: S_0}{} 
&\hspace*{1cm}& 
\Rule{}{}{\Gamma\vdash x :T}{$x:T\in\Gamma$}
\\ \ \\
\Rule{}{}{\Gamma\vdash K : K_0}{}
&&
\Rule{}{\Gamma\vdash M : T \quad \Gamma\vdash N : U}{\Gamma\vdash MN : V}{$T (U) = V.$}
\end{array}
\]

\caption{The Summary Type System} 
\label{fig:summary}
\end{figure}

\begin{theorem}[unique\_types]
If $\Gamma \vdash M : T_1$ and $\Gamma \vdash M : T_2$ then $T_1 = T_2$. 
\end{theorem}

\begin{theorem}[reduction\_preserves\_typing]
If there is a type derivation $\Gamma\vdash t : T$ and $t$ reduces to $t_2$ then there is a derivation $\Gamma\vdash t_2 : T$. 
\end{theorem}

\section{Type Inference}
\label{sec:infer}

Since types are unique, if they exist, it is routine to produce a type inference algorithm {\bf infer}. This has been formalised in Rocq and implemented in OCaml to further assess the practicality of our approach. The cases relevant to combinatory types are given in Figure~\ref{fig:infer}, with {\tt << \ldots  >>} used to indicate where the cases for abstract types have been elided. Function {\bf infer\_app} implements type application.
The algorithm is able to find types whenever they exist.

\begin{theorem}
$\Gamma \vdash M : T$ if and only if\  ${\bf infer}~\Gamma~M = T$. 
\end{theorem}

\lstdefinelanguage{ocaml}{ 
    mathescape=true,
    texcl=false, 
    morekeywords=[1]{let, rec, match, with, in, bind, when},
    morekeywords=[2]{true, false},
    morekeywords=[3]{Map, Set, option},
    morecomment=[s]{(*}{*)},
    morecomment=[s]{<<}{>>},
    showstringspaces=false,
    morestring=[b]",
    morestring=[d]’,
    tabsize=2,
    extendedchars=false,
    sensitive=true,
    breaklines=false,
    basicstyle=\small,
    captionpos=b,
    columns=[l]flexible,
    identifierstyle={\ttfamily\color{black}},
    keywordstyle=[1]{\ttfamily\color{dkblue}},
    keywordstyle=[2]{\ttfamily\color{dkblue}},
    keywordstyle=[3]{\ttfamily\color{dkblue}},
    stringstyle=\ttfamily,
    commentstyle={\ttfamily\color{dkgreen}},
    literate=
    {->}{{$\rightarrow\;$}}1
    %
}[keywords,comments,strings]

\begin{figure}
\begin{lstlisting}[language=ocaml]
let rec infer_app ty vty =
  match ty with
  (* combinatory types *)
  | Kty -> Some (K1ty vty)
  | K1ty uty -> Some uty
  | Sty -> Some (S1ty vty)
  | S1ty uty -> (
      let rty = S2ty (uty, vty) in
      (* unpack_constructor rty = Some (fty, tty) if and only if
         tagged_ty{fty, tty} = rty *)
      match unpack_constructor rty with
      | None -> Some rty (* not the form of a constructor *)
      | Some (fty, tty) -> (* may be a constructor *)
        << special cases for introducing abstract types have been elided >>
    )
  | S2ty (uty, wty) ->
      let%bind.Option wvty = infer_app wty vty in
      let%bind.Option uvty = infer_app uty vty in
      infer_app uvty wvty
  << special cases for eliminating abstract types have been elided >>
  | _ -> None


let rec infer gamma m =
  match m with
  | Ref x -> get x gamma
  | Sop -> Some Sty
  | Kop -> Some Kty
  | App (m1, m2) -> (
      match (infer gamma m1, infer gamma m2) with
      | Some ty, Some vty -> infer_app ty vty
      | _, _ -> None)
\end{lstlisting}
\caption{Type Inference Cases for Combinatory Types in OCaml} 
\label{fig:infer}
\end{figure}

Also, it will usually fail if the term has no type. In some exceptional cases, the algorithm will not terminate. For example, consider the self-application of $(SII)$. Although $SII$ has type $S_2 I_0 I_0$ the self-application of this type is not defined, and the inference algorithm goes into a loop. In practice, it is hard to induce such behaviour, since recursive functions defined by $Z$ are handled using special cases. If required, non-termination can be avoided by imposing a bound on the number of calls to inference. Indeed, the Rocq implementation limits the depth of recursive calls to ensure termination.

The OCaml implementation \cite{BaderCombTypes2025} instruments the function {\bf infer} with a counter that can be used to either measure or limit the total number of type application steps, i.e. calls to {\bf infer\_app}. If the limit is reached, inference fails artificially.
Note that this is different from the Rocq approach of limiting recursion depth. Counting calls instead is a better proxy for the total running time of inference, which we assess here.

Table~\ref{table:fuel} shows a set of concrete experiments that cover all the functions and data types introduced in this paper, a primitive recursive addition {\bf plus} and more.
The measurements suggest that type inference terminates, for realistic programs and common type errors, after a number of {\bf infer\_app} calls that is linear in the size of the term (number of operators).
In particular, the ratio between calls and size stays below 2 for most, below 3 for all tested examples.

\begin{table}
\small
\begin{center}
\begin{tabular}{|l|l|l|l|l|}
\hline
term & has type & size & \#calls & \#calls / size \\
\hline
{\bf cond tt ff tt} & yes & 149 & 244 & 1.64 \\
{\bf cond tt tt zero} & no & 147 & 242 & 1.65 \\
{\bf cond} ({\bf cond tt ff tt}) {\bf ff tt} & yes & 279 & 470 & 1.68 \\
{\bf pair ff tt} & yes & 97 & 147 & 1.52 \\
{\bf snd (pair ff tt)} & yes & 106 & 167 & 1.58 \\
{\bf pair} ({\bf pair ff tt}) {\bf tt} & yes & 172 & 273 & 1.59 \\
{\bf successor tt} & no & 57 & 74 & 1.30 \\
{\bf successor zero} & yes & 58 & 75 & 1.29 \\
{\bf successor$^3$ zero} & yes & 134 & 187 & 1.40 \\
{\bf successor$^{1000}$ zero} & yes & 38020 & 56019 & 1.47 \\
{\bf predecessor zero} & yes & 104 & 164 & 1.58 \\
{\bf isZero} & yes & 103 & 153 & 1.49 \\
{\bf isZero zero} & yes & 123 & 180 & 1.46 \\
{\bf isZero} ({\bf successor zero}) & yes & 161 & 236 & 1.47 \\
{\bf isZero tt}  & no & 122 & 178 & 1.46 \\
{\bf case} ({\bf pair isZero $I$}) ({\bf inl} ({\bf pair zero tt})) & yes & 523 & 954 & 1.82 \\
{\bf case} ({\bf pair isZero $I$}) ({\bf inr} ({\bf pair zero tt})) & yes & 526 & 957 & 1.82 \\
{\bf case} ({\bf pair $I$ $I$}) ({\bf inr} ({\bf pair zero tt})) & no & 426 & 802 & 1.88 \\
{\bf lam} $x$ ({\bf isZero} $x$) {\bf zero} & yes & 140 & 204 & 1.46 \\
{\bf lam $x$ $x$ zero zero} & yes & 55 & 58 & 1.05 \\
{\bf lam $x$ $x$ zero tt} & no & 54 & 57 & 1.06 \\
{\bf cond\_mono}\{{\bf zero}\} & yes & 207 & 335 & 1.62 \\
{\bf cond\_mono}\{{\bf zero}\} {\bf tt} & yes & 226 & 354 & 1.57 \\
{\bf cond\_mono}\{{\bf zero}\} {\bf tt zero} & yes & 246 & 374 & 1.52 \\
{\bf cond\_mono}\{{\bf zero}\} {\bf tt ff} & no & 248 & 376 & 1.52 \\
{\bf plus} ({\bf successor zero}) & yes & 954 & 1130 & 1.18 \\
{\bf plus} ({\bf successor zero}) {\bf zero} & yes & 974 & 1367 & 1.40 \\
{\bf plus} ({\bf successor zero}) {\bf tt} & no & 973 & 1311 & 1.35 \\
{\bf cons} ({\bf pair ff} ({\bf nil zero})) & no & 188 & 294 & 1.56 \\
{\bf cons} ({\bf pair ff} ({\bf nil tt})) & yes & 187 & 293 & 1.57 \\
{\bf cons} ({\bf pair ff} ({\bf cons} ({\bf pair tt} ({\bf nil tt})))) & yes & 321 & 519 & 1.62 \\
{\bf fold\_left}\{{\bf plus}\} & yes & 1638 & 1670 & 1.02 \\
compiler of a toy language & yes & 48016989 & 119811071 & 2.50 \\
compiler of a toy language (subroutine) & yes & 22563 & 56888 & 2.52 \\
$S^4$ (= $S~S~S~S$) & yes & 5 & 10 & 2.00 \\
$S^{10}$ & yes & 11 & 70 & 6.36 \\
$S^{100}$ & yes & 101 & 7450 & 73.76 \\
$(SII)(SII)$ & no  & 14 & $\infty$ & $\infty$ \\
\hline
\end{tabular}
\end{center}
\caption{Selection of Metrics about Calling {\bf infer} on Various Terms}
\label{table:fuel}
\end{table}

A notable, albeit unrealistic, counterexample is the repeated application of $S$ to itself. In this case we have $|S S S S S S \ldots | = S_2 S_1 (S_2 S_1 (S_2 S_1 \ldots ))$, where term and type have the same size. Adding a single $S$ requires computing {\bf infer\_app} $|S S S S S S \ldots|~|S|$, which recursively descends into and rewrites $S_2 S_1 (S_2 S_1 (S_2 S_1 \ldots ))$ using the {\bf S2ty} branch in Figure~\ref{fig:infer}. Since this is linear in the size of the term, the total work for $k$ operators is $\mathcal{O}(k^2)$.
Apart from degenerate cases like this or $(SII)(SII)$, the algorithm terminates quickly, regardless of whether the term has a type or not. For example, inference on both {\bf successor zero} and {\bf successor tt} has a ratio of about 1.3.

We further tested inference against a small compiler for a toy language
that uses recursion and a plethora of Scott-encoded data types (numbers, strings, option types, parser state, tokens, ASTs, etc) but is completely oblivious to our data types and tagging approach. As a result, {\bf infer\_app} wrangles nothing but combinatory types, which we believe represents a realistic worst-case scenario.
The ratio turns out to be merely 2.5, while the largest ratio across all subroutines of the compiler barely differs, at 2.52.

We therefore recommend limiting the total number of calls to at least 3 times the input term's size. Larger factors will be able to accommodate increasingly unlikely kinds of terms.
It is worth noting that even at a factor of (say) 100, inference rejects the term $(SII)(SII)$ swiftly.
We conclude that inference on realistic programs is practical, as it even terminates on programs oblivious to our tagging in linear time, with some small factor.

\section{Related Work}
\label{sec:related}
The type system introduced in this paper diverges
significantly from traditional approaches to polymorphism. 
There is very little literature on type systems for combinatory logic per se. Aside from being mentioned by Curry et al \cite{curry1972combinatory}, almost all work has interpreted the combinators as $\l$-abstractions within a typed $\l$-calculus, to which we now turn.

Unlike System~HM \cite{mil78} and System~F \cite{GLT89}, our framework
avoids both type abstraction and application, which simplifies
both the type syntax and inference algorithm. This simplification
requires some use of dummy values as placeholders for types, but enhances transparency for
static analysis. 

{\em Intersection types}, pioneered by Coppo et al \cite{cdv80}, 
allow multiple types to be ascribed to the same term, supporting a
refined analysis of program behaviour. However, such systems generally
lack principal types and often require complex inference mechanisms
\cite{barendregt2013lambda}. Our system avoids these complications by
ensuring that each combinator has at most one type, relying instead on
the type structure to encode information otherwise
captured by intersections.

{\em Semantic subtyping}, as in the work of Frisch et al \cite{frisch2008semantic}, supports a rich subtyping
relation based on set-theoretic semantics of types. However, set-theoretic concepts, such as unions and complements, make both reasoning and implementation difficult. Here, information hiding is achieved by the transition from combinatory types to abstract types, which is much simpler because it is structural not semantic. 

Other systems, such as those based on {\em refinement types} \cite{freeman1991refinement,rondon2008liquid}  enhance expressiveness by annotating types with logical predicates, often requiring SMT solvers for type checking. Our system is simpler because it is structural, not reliant on external logical mechanisms.

\section{Future Work}
\label{sec:future}

\subsection{Impredicativity}

Although type inference for System~F is known to be undecidable \cite{WellsJB:typtcs}, some impredicative types have been inferred. For example,  Serrano et al  \cite{10.1145/3408971} show how to type the composition of a function $f: (\forall a. {\bf List}\{a\} \to {\bf List}\{a\}) \to {\bf Nat} $ and $g: {\bf Bool}\to \forall a. 
{\bf List}\{a\} \to {\bf List}\{a\}$ to get a function of type ${\bf Bool} \to {\bf Nat}$. Using combinatory types, we could approach this challenge in two ways. One is to replace $\forall a. {\bf List}\{a\} \to {\bf List}\{a\}$ by some abstract type of polymorphic functions. The other would be to develop an abstract type for quantification in general.

\subsection{Typing Modules} 

By module, we mean some useful collection of types and terms equipped with an interface that determines how it may be used, as in \cite{10.1145/174675.176927}. 
 For example, a module for lists might include a polymorphic data type of lists, plus functions for folding over lists, etc. As long as the interface is respected, the implementation of the module's contents can be maintained, re-implemented, etc.  It seems plausible that modules can be given abstract types, and so become first-class terms of a combinatory calculus.

\subsection{Static Analysis}

Although type inference is a small part of static program analysis, it suggests that combinatory type systems may be a simpler setting in which to work.
Traditionally, the interesting recursive programs did not have normal forms, which makes analysis difficult or impractical. So static analysis works with syntax trees that represent these functions, which separates the syntax from the semantics. However, in the combinatory approach, all programs are normal forms so there is no need to separate syntax from semantics. 
The combinatory types take advantage of this, to support a refined program analysis that is able to reveal complete information about the program structure, while relying on abstract types to hide this information as necessary. That is, the general machinery need not obscure the structures that are central to the analysis, such as Min\'e's numerical domains \cite{mine2006octagon}  or Sui's  inter-procedural analyses  \cite{sui2016svf}.

\subsection{Tree Calculus} 

Tree calculus \cite{tcb} goes beyond the {\em extensionality} of combinatory logic to support {\em intensionality}, including the ability to decide equality of programs, to reveal the label of a tagged term, and to support a self-interpreter that acts on itself directly, without encoding to a syntax tree. It supports all program analyses to the same extent that combinatory logic supports all numerical analyses. 
%
%
Recent work \cite{10.1145/3704253.3706138} has shown how to type tree calculus using combinations of quantified function types that are not found in HM. The combinatory type system is a stepping stone to combinational type system for tree calculus that will support type inference for typed program analyses.
%
That done, it will be natural to internalise the types and type-level computations as terms and intensional programs, and then blend them in a system of dependent types.

\section{Conclusions}
\label{sec:conclusions}

Combinatory logic provides better support for static program analysis than $\l$-calculus.  The principal reason for this is that all Turing-computable functions can be represented by normal forms, so that we can identify these with the programs, and analyse them statically. There is no need to encode programs as syntax trees before analysing them. Unlike $\l$-calculus, there is no need to modify the term syntax to increase expressive power. For example, data type constructors can be represented by combinators that are not used to represent $\l$-abstractions.  Then they can be typed by abstract types built upon combinatory types. Such types are simple, in that there are no type variables or quantification. Even function types can be represented as abstract types. Yet they support polymorphism which is revealed during application. This is enough to support the usual polymorphism of the Hindley-Milner type system, and more, such as tuples of polymorphic programs, and perhaps even program modules.  It has an effective type inference algorithm. More generally, it provides a simple, yet powerful system in which to do static program analysis. 


\end{document}